\documentclass[a4paper,11pt]{article}
\usepackage{pos}
\usepackage{amsmath,amsfonts,latexsym,hhline,stmaryrd,color,verbatim,graphicx,epstopdf,multirow}
\usepackage{slashed}
\usepackage{lineno}
\usepackage{color}
\usepackage[utf8]{inputenc}
\usepackage{booktabs}
\usepackage{subfig}
\usepackage{pifont}
\usepackage{url}

\title{Searching for Dark Matter with circularly polarised photons from the Galactic Centre}

\ShortTitle{Searching for DM with circularly polarised photons from the GC}

\author*{Marina Cermeño}
\author{Céline Degrande}
\author{Luca Mantani}

\affiliation{Centre for Cosmology, Particle Physics and Phenomenology (CP3), Universite Catholique de Louvain,\\
Chem. du Cyclotron 2, 1348,  Louvain-la-neuve, Belgium}

\emailAdd{marina.cermeno@uclouvain.be}
\emailAdd{celine.degrande@uclouvain.be}
\emailAdd{luca.mantani@uclouvain.be}

\abstract{
Conventional indirect dark matter (DM) searches look for an excess in the electromagnetic emission from the sky that cannot be attributed to known astrophysical sources, but its polarisation has not been explored to date.

In this proceeding, we argue that photon polarisation is an important feature to understand new physics interactions. In particular, circular polarisation can be generated from Beyond the Standard Model interactions if they violate parity and there is an asymmetry in the number density of the initial state particles which participate in the interaction. We consider a simplified model for fermionic (Majorana) DM and study the circularly polarised gamma rays below 10 GeV from DM cosmic ray electron interactions. We study the differential flux of positive and negative polarised photons from the Galactic Centre and show that the degree of circular polarization can reach up to 90\%. Finally, we discuss the detection prospects of this signal in future experiments.
}

\FullConference{
European Physical Society Conference on High Energy Physics - EPS-HEP2021 -\\
			26-30 July, 2021\\
			Online}

\begin{document}
\maketitle
\section{Introduction}
The existence of dark matter (DM) is strongly supported by several evidence that has been accumulated over the last decades~\cite{zwicky1, zwicky2, vera, bulletcluster1, bulletcluster2, bc3, cmb, cmb1, cmb2}. Despite the world-wide experimental effort, the identity of DM is still unknown and the need for novel ideas is more pressing than ever.

Under the assumption that DM is made of beyond Standard Model (BSM) particles, several experiments have been designed. In particular, one of the most promising strategy is to indirectly detect DM from high density regions, e.g. the Galactic Center (GC)~\cite{vanEldik:2015qla}, where it can either decay or annihilate producing radiation that reaches our detectors on Earth. The main complication of this strategy is the fact that it might be non trivial to distinguish a signal produced by DM from the background produced by other astrophysical sources.

A distinguishable evidence would come from  the observation of peaks in the astrophysical spectrum, such as the ones predicted in~\cite{Bringmann:2012vr, Garny:2013, Kopp:2014, Okada:2014zja, Garny:2015wea, Kumar:2016cum, Bartels:2017dpb} due to the annihilation of fermionic Majorana DM. In this scenario, the 2 to 2 annihilation of DM via a charged mediator is p-wave suppressed being internal bremsstrahlung more likely to happen. From this process a sharp peak in the spectrum is expected at the DM mass energy scale for a mediator close in mass to the DM particle. While DM coupling to quarks is already tightly constrained for this scenario~\cite{Garny:2015wea}, the leptophilic case still offers a wide window to explain the DM nature.
In this context, not only annihilation will provide a peak in the spectrum, for this class of models another peak is expected at the DM-mediator mass splitting energy due to the scattering of DM with cosmic ray (CR) electrons, as it has been analysed in~\cite{Gorchtein:2010xa, Profumo:2011jt, Huang:2011dg, Gomez:2013qra}. Depending on the parameters of the model (mass splitting, coupling, etc) the number of photons at this energy can be comparable with the one due to annihilation. Nevertheless, what can certainly differentiate the two signals is the fact that only the flux of photons coming from the DM-CR electron scattering can be circularly polarised.

A net circular polarisation signal can be observed in the sky when when there is an excess of
one circular polarisation state over the other. Since photons flip helicity under parity, parity must be violated in at least one of the dominant photon emission processes. But parity 
(P) violation is not the only condition required, there must be either an asymmetry in the number density of one of the particles in the initial
state or CP must be violated by the interactions at play. Therefore, an interaction where the initial state is a CP-eigenstate, such as the annihilation of a Majorana particle, cannot generate circular polarisation even if it violates parity.
Using this arguments, in~\cite{Kumar:2016cum,Elagin:2017cgu,Gorbunov:2016zxf,Boehm:2017nrl, Huang:2019ikw, Balaji:2019fxd, Balaji:2020oig} it is suggested that DM and neutrinos can generate circular polarised signals in X-rays or gamma-rays through decays and interactions  with SM particles, pointing out an unexplored way
to look for new physics. In particular, a net circular polarisation signal can be generated by DM interactions with ambient CRs. Motivated by this, in our work we perform for the first time the full computation of the flux of circularly polarised photons coming from the interactions between a Majorana DM fermion and CR electrons in the GC. After finding that the circular polarisation fraction can reach values higher than $90\%$, we argue that this feature and their energy dependence could be used to reveal these BSM interactions as well as to study their nature.

\section{Circular polarisation}

In this section we present the formalism used for the photon polarisation states. As previously anticipated, in order to have a source of net polarisation both P and CP symmetries need to be violated by the underlying physics process.

Photon helicities flip under P transformations. Therefore, if the physics is invariant under parity it means that no helicity is preferred to the other, as they both couple in the same way. The simplest way to have a parity violating interaction is to have chiral interactions, with right and left-handed fermion components coupled in different manners.

However, if P is violated but CP is not, the CP conjugate process will generate the opposite helicity with the same rate, making it effectively impossible to produce a net circular polarisation. In this work we consider a scenario in which CP is violated not at the fundamental level, but by means of an asymmetry in one of the particles of the initial state, i.e. the number density of the particle is not the same as the number density of its antiparticle. In this situation, the CP process cannot counter-balance the production of polarised photons.

Four our calculations, we use the same convention for the photon polarisation vectors that is used in~\cite{Boehm:2017nrl}. We consider a photon with momentum $k^\mu=(k_0, k_x, k_y, k_z)$, whose two possible transverse polarisation vectors are
\begin{equation}
\epsilon_1^\mu(k)=\frac{1}{k_0 k_T}(0, k_x k_z, k_y k_z, -k_T^2)    
\end{equation}
and
\begin{equation}
\epsilon_2^\mu(k)=\frac{1}{k_T}(0, -k_y, k_x, 0), 
\end{equation}
where $k_T=\sqrt{k_x^2+k_y^2}$. 
The positive and negative photon circular polarisation vectors are then defined as
\begin{equation}
\epsilon_{\pm}^\mu(k)= \frac{1}{2}(\mp\epsilon_1^\mu(k)-i \epsilon_2^\mu(k) ).   
\end{equation}

These definitions allow us to define the squared helicity amplitudes $\mathcal{A}_-= \sum_{spins} |\epsilon^{\mu}_{-}\mathcal{M}_\mu|^2$ and $\mathcal{A}_+= \sum_{spins} |\epsilon^{\mu}_{+}\mathcal{M}_\mu|^2$ such that their sum is equivalent to the total averaged amplitude. For a P violating interaction $\mathcal{A}_- \neq \mathcal{A}_+$ and a net circularly polarised spectrum is expected.

\section{Model}

\begin{figure}[t!]
  \centering
  \includegraphics[scale=0.23]{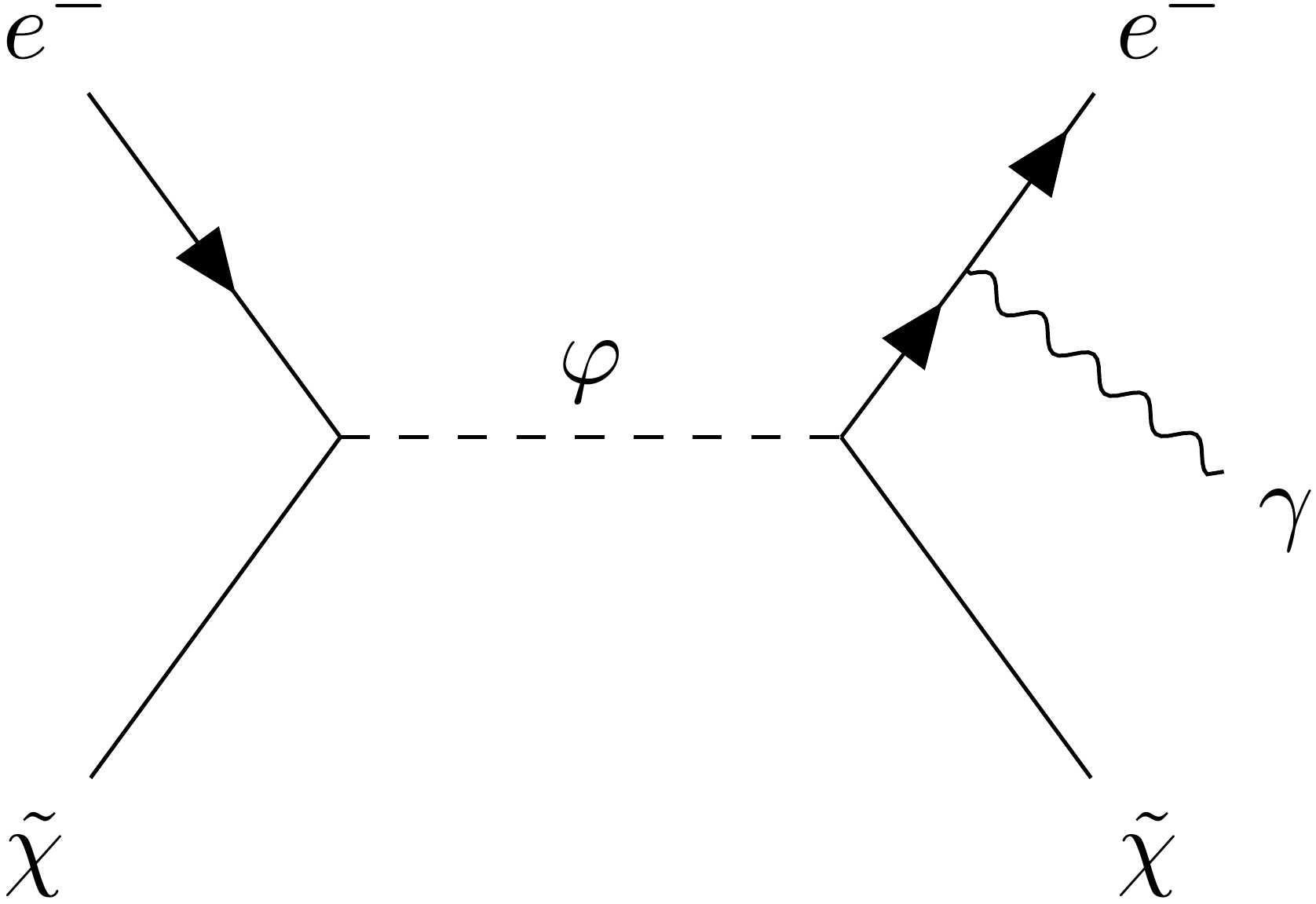} \qquad
  \includegraphics[scale=0.23]{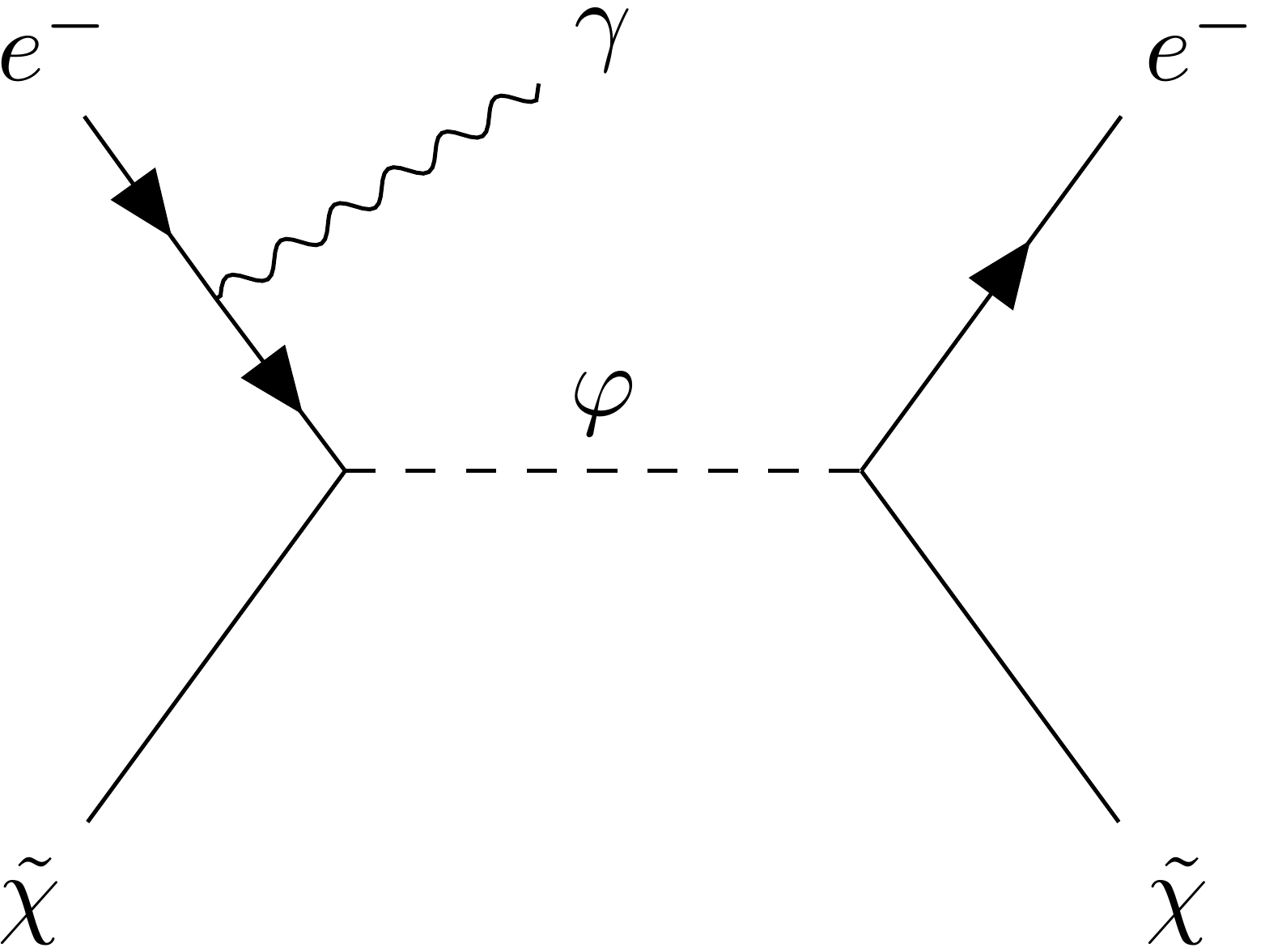} \qquad
  \includegraphics[scale=0.23]{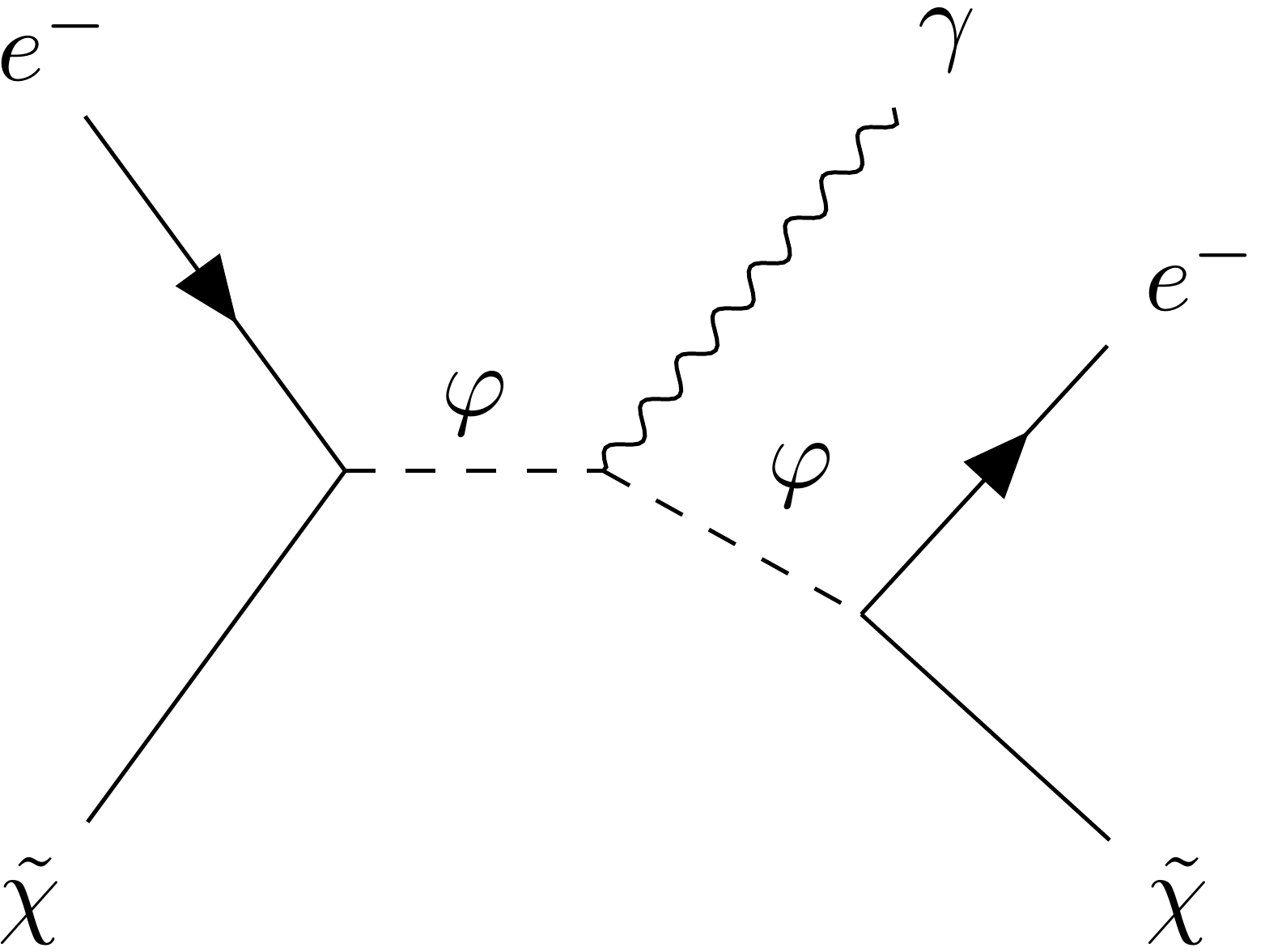}
  \caption{
\label{fig:diagrams}
Diagrams for the resonantly enhanced 2 to 3 radiative processes of electrons scattering with a Majorana DM particle through a right-handed scalar mediator. }
\end{figure}

In order to compute the circular polarisation asymmetries from the expected flux of photons coming from DM-CR scattering in the GC,
following~\cite{Profumo:2011jt, Kopp:2014, Garny:2015wea}, we consider a t-channel simplified model in which a Majorana DM candidate $\tilde{\chi}$ is coupled to right-handed electrons by means of a charged scalar mediator $\varphi$. The SM Lagrangian is therefore extended with two degrees of freedom and the dark sector interactions are given by

\begin{equation}
    \mathcal{L}_{DM} = i \bar{\psi}_{\tilde{\chi}}(\slashed{D} - m_{\tilde{\chi}})\psi_{\tilde{\chi}} + D_\mu \varphi^\dagger D^\mu \varphi - m_\varphi \varphi^\dagger \varphi + ( a_R \, \bar{e}_R \, \psi_{\tilde{\chi}} \, \varphi + h.c.) \, .  
\end{equation}

This simplified model can be interpreted in a supersymmetry (SUSY) context where $\tilde{\chi}$ is the lightest neutralino and $\varphi$ is the right-handed selectron. The scalar mediator is carrying the same quantum numbers of the right-handed electron, since the DM is a singlet of the SM gauge groups. The model parameter space is three dimensional, i.e. is characterised uniquely by the set of parameters $\{m_{\tilde{\chi}}, m_\varphi, a_R \}$. Additionally, in order to ensure DM stability, the mass of the mediator has to be bigger than the mass of the DM candidate ($m_\varphi > m_{\tilde{\chi}}$). By considering a coupling solely to the right-handed component of the electron, we maximise the parity violation of the model and consequently the possibility to produce a net signal of circularly polarised photons.

The choice of a t-channel model is motivated by the fact that they can lead to a kinematic enhancement in the DM-CR scattering by means of a resonant contribution when the mass splitting $\Delta M = m_\varphi - m_{\tilde{\chi}}$ is of the order of the CR energies ($\sim$ few GeV). Thanks to this feature, one can exploit the resonance to probe the degenerate region of the parameter space which is difficult to access at colliders and other experiments. The relevant diagrams of the DM-CR interactions which are resonantly enhanced can be seen in Fig.~\ref{fig:diagrams}.

\section{Polarised photon flux from cosmic ray scattering}

In this section we present results for the photon spectrum generated by DM-CR scattering. In particular, we focus on the GC, due to its high DM density and large electron flux. The leading contribution to the flux comes from the $2$ to $3$ scattering $\tilde{\chi} e^- \rightarrow \tilde{\chi} e^- \gamma$.

We define the flux of circularly polarised photons at a distance $r_\odot$ from the GC as
\begin{eqnarray}
\frac{d\Phi_{e\tilde{\chi}, pol}}{dE_\gamma}=\bar{J}\frac{1}{m_{\tilde{\chi}}}\int d\Omega_\gamma \int dE_e \frac{d\phi}{dE_e} \left|\frac{d^2\sigma_+}{d\Omega_\gamma dE_\gamma}(E_e,\theta_\gamma,E_\gamma )-\frac{d^2\sigma_-}{d\Omega_\gamma dE_\gamma}(E_e,\theta_\gamma,E_\gamma ) \right|,
\label{polflux1}
\end{eqnarray}
where $\Omega_\gamma$ is the solid angle between the emitted photon and the incoming CR electron (with $\theta_\gamma$ the polar coordinate), and $E_e$ and $E_\gamma$ are the incoming electron and the outgoing photon energies. The $+$ and $-$ signs in the differential cross section indicate the positive and negative circular polarisations. The CR electron energy spectrum, which has a relevant impact on the overall flux and on the degree of net polarisation as well, is described by $\frac{d\phi}{dE_e}$. The spatial dependence of the DM and electron distributions are taken into account in the factor
\begin{eqnarray}
\bar{J}(\Delta \Omega_{\rm obs})=\frac{1}{\Delta \Omega_{\rm obs}}\int_{\Delta \Omega_{\rm obs}} d\Omega \int_{\rm l.o.s} ds \; \rho(r(s,\theta)) \; f(r(s, \theta)),
\label{jbar}
\end{eqnarray}
which integrates over the line of sight and solid angle of observation of the experiment, $\Omega_{\rm obs}$, the product of the DM density distribution, $\rho(r)$, and the function
$f(r)=\frac{e^{-\frac{r}{r_0}}}{e^{-\frac{r_\odot}{r_0}}}$,
 which takes into account the fact that the CR flux is larger in the GC vicinity~\cite{Profumo:2011jt,Strong:2004td}. Here,  $r_\odot= 8.5$ kpc is the distance from the GC to the Earth and $r_0=4 \; \rm kpc$.

Being the cross section of the process dominated by the resonant contributions, especially for small mass splittings, we are in the ideal scenario to employ the narrow width approximation (NWA)~\cite{Berdine:2007uv}. A more thorough description of the performed calculation can be found in Ref.~\cite{Cermeno:2021rtk}. Thanks to the resonant enhancement, the cross section is not proportional to $a_R^4$ as naively expected, but to $a_R^2$.

\subsection{Results}
\label{subsec:results}

\begin{figure}[t!]
  \centering
  \includegraphics[width=.45\columnwidth]{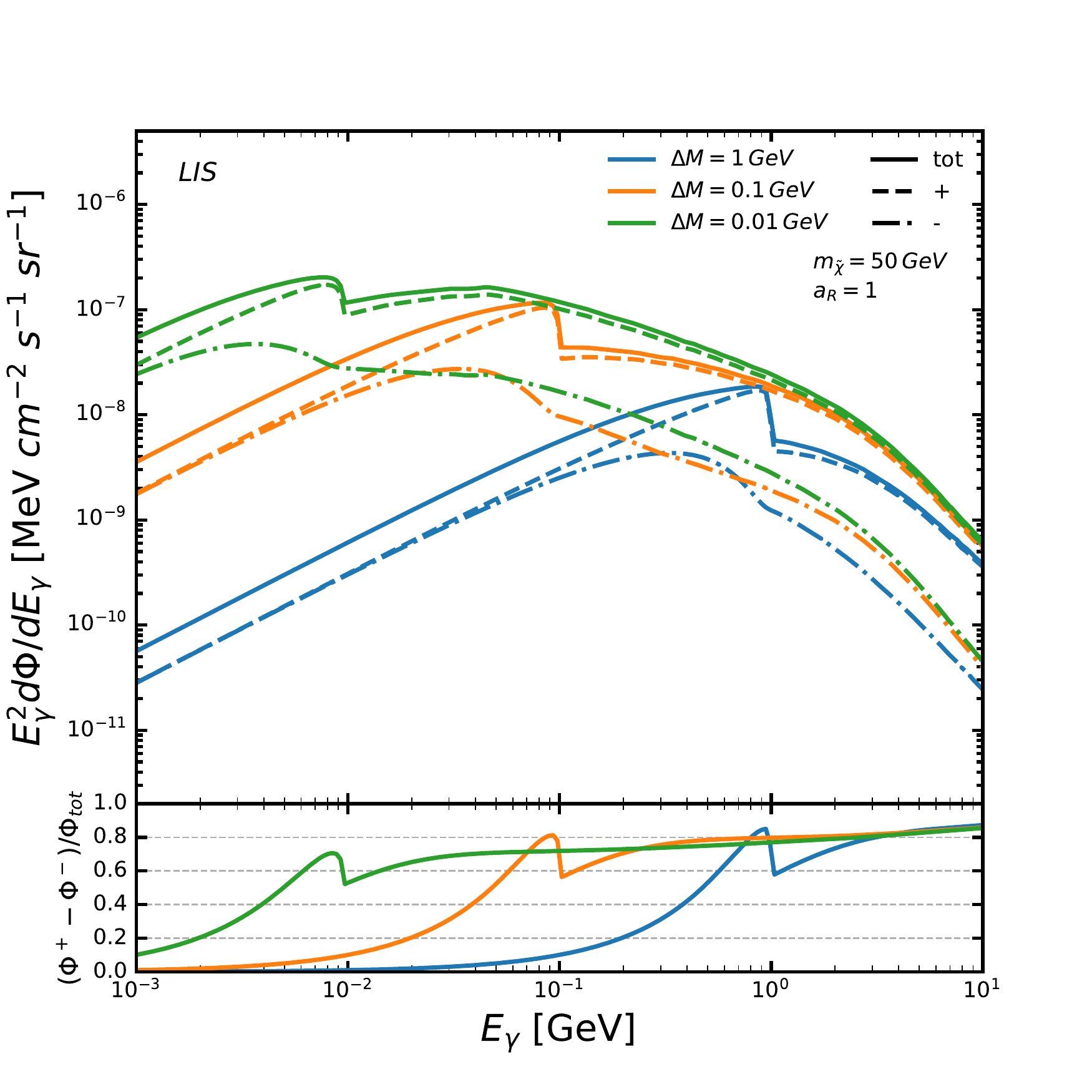}
  \includegraphics[width=.45\columnwidth]{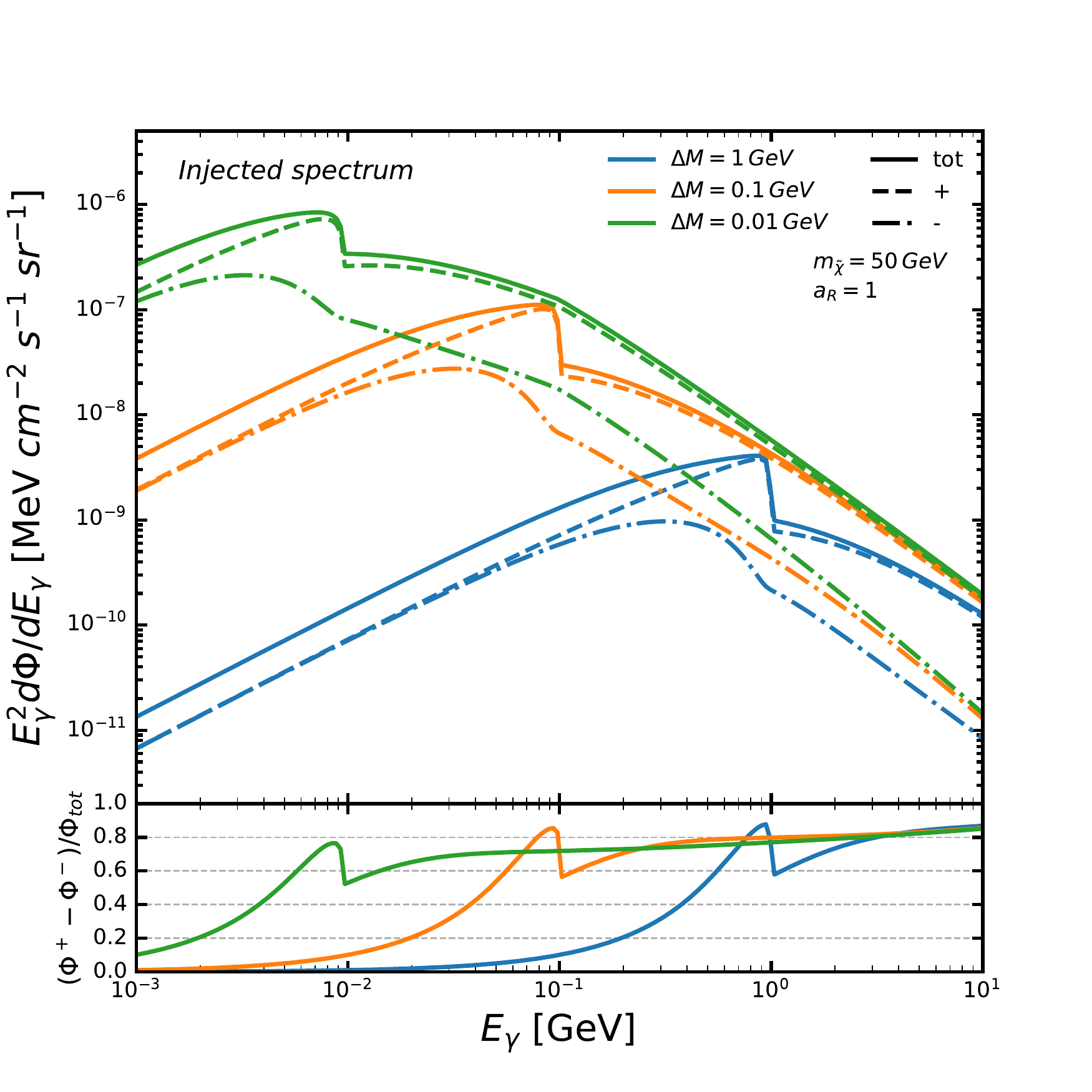}\\
  
  \includegraphics[width=.45\columnwidth]{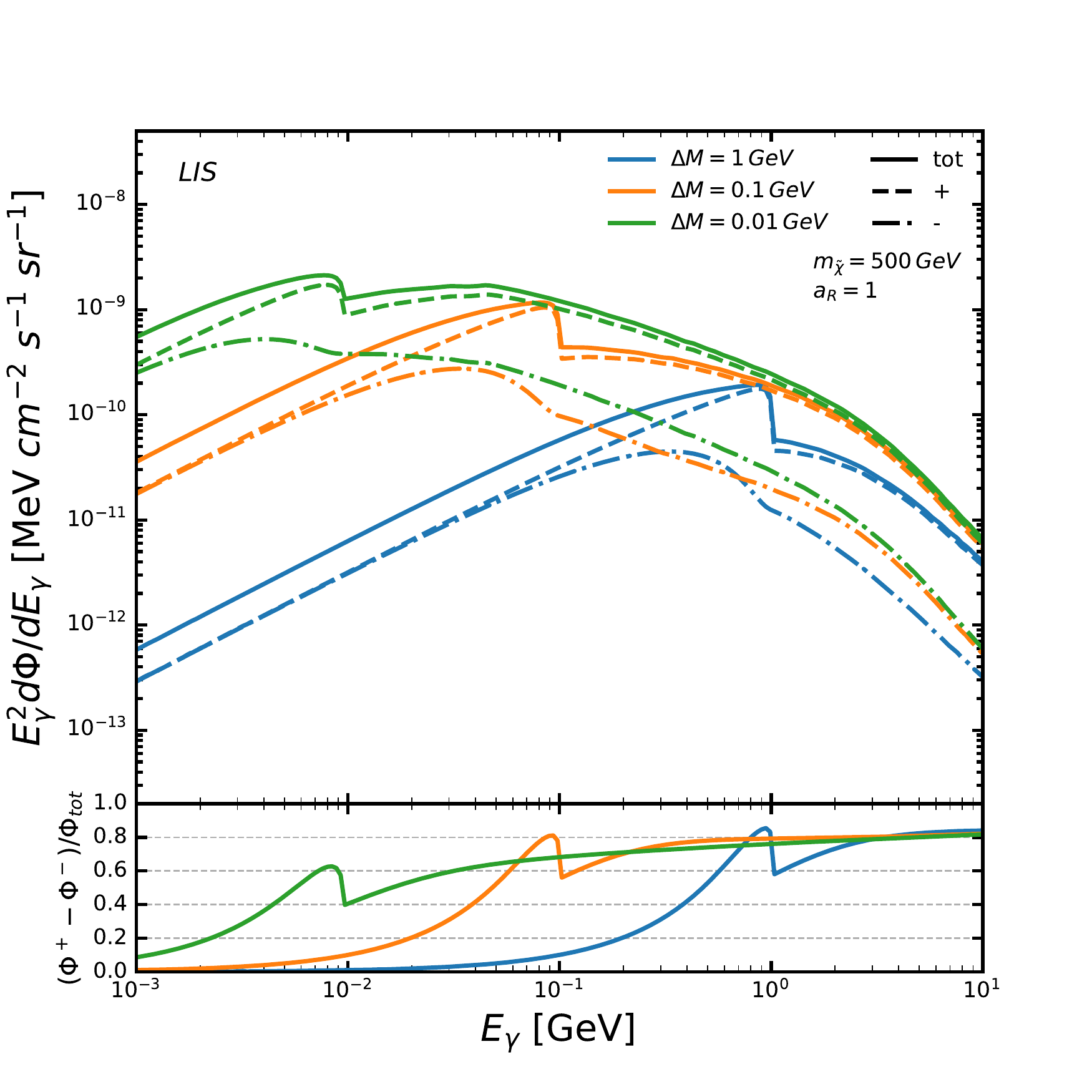}
  \includegraphics[width=.45\columnwidth]{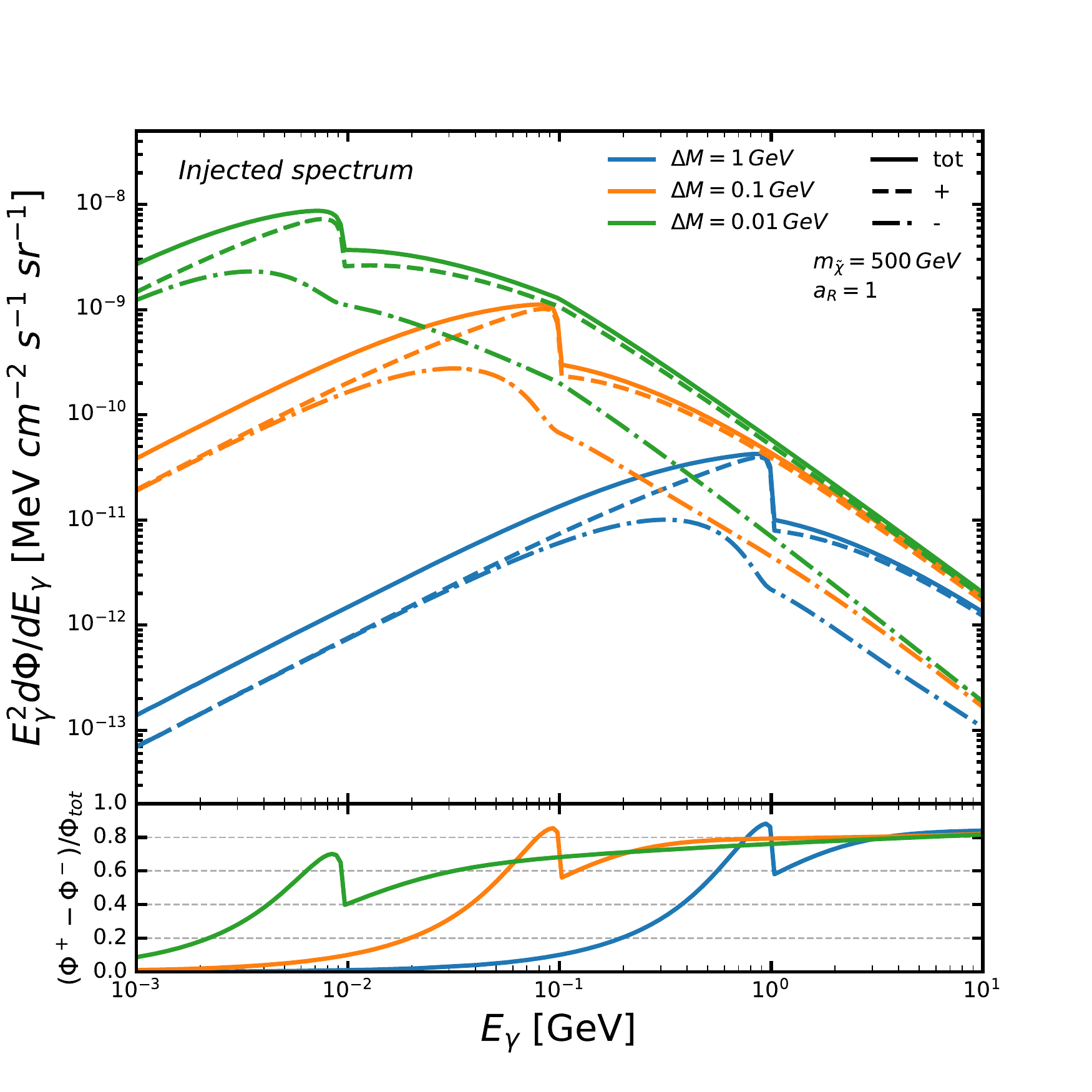}
  \caption{
    \label{fig:flux}
  The flux spectrum of negative and positive circularly polarised photons in dash-dotted and dashed lines, respectively, and the sum of them, in solid lines, from the scattering of CR electrons off DM. In the lower part of each plot the difference between the flux of positive and negative polarised photons over the total one, i.e the circular polarisation asymmetry $A_{pol}$, can be seen.}
\end{figure}

In the following we present the photon fluxes and the circular polarisation asymmetries expected from the scattering of DM and CR electrons in the GC. Regarding the DM density, we assume the Einasto profile which provides a $\bar{J}_{\rm Ein}=1.67 \; 10^{24}\; \rm GeV/ cm^2$ for the typical angular resolution of Fermi-LAT~\cite{Fermi:2009} and e-ASTROGAM~\cite{DeAngelis:2017gra}, i.e., $\theta_{\rm obs}= 1^{\circ}$, which corresponds to $\Delta \Omega_{\rm obs} \sim 10^{-3}$. 

In Fig.~\ref{fig:flux} we show the flux spectrum of negative and positive circularly polarised photons and their sum (dash-dotted, dashed and solid lines respectively). We fix the DM mass to $m_{\tilde{\chi}}=50$ GeV (upper panel) and $m_{\tilde{\chi}}=500$ GeV (bottom panel) and take different values of $\Delta M$, namely $1$, $0.1$ and $0.01$ GeV. In the lower part of each plot the difference between the flux of positive and negative polarised photons over the total one, i.e. the circular polarisation asymmetry $\mathcal{A}_{pol}$, is reported.
As naively expected, we observe that the higher the DM mass and $\Delta M$ are, the lower the flux is.

With respect to the electron spectrum, in the left panel of Fig.~\ref{fig:flux} we show results considering the local interstellar spectrum, extracted from Fig. 9 of~\cite{Vittino:2019yme}.
On the right panel of Fig.~\ref{fig:flux} instead, we can see the results that we get if we consider the electron injected spectrum. In order to be consistent with the local interstellar spectrum interpolated from data of~\cite{Vittino:2019yme}, we use the injected spectrum of the $1$ break model from this Reference.

We observe that the asymmetry of polarised photons coming from these interactions can reach up to $90\%$. For a fixed value of $\Delta M$, the peak of the flux is obtained at $E_\gamma=\Delta M$, which is also where the polarisation asymmetry displays a maximum. The detection of this kind of signature is complementary to the observation of a peak at $E_\gamma \sim m_{\tilde{\chi}}$ coming from self-annihilation of DM, which cannot produce a net flux of circular polarised photons (initial state is CP even). While from the DM-CR scattering one can gain information on the mass splitting $\Delta M$, from the annihilation one can learn about the mass of DM. In this perspective, the circular polarisation could be used as a characterisation feature to pinpoint the nature of DM.
However, it must be noticed that, taking into account the background of photons coming from the GC in this energy region predicted by~\cite{Gaggero:2017dpb} (Fig. 2 black solid line), even if we consider the highest flux of photons found for the injected spectrum, we would need an exposure time of $\Delta t \sim 10^{10}$~s for a $3 \sigma$ evidence. Nowadays the only efficient techniques to measure photon circular polarisation in this energy range band are based on the measurement of the secondary asymmetry of photons via Compton scattering \cite{Elagin:2017cgu}. Therefore, unless a new technique which exploits the polarisation fractions with the objective of increasing sensitivity is developed, these signals will not be able to be detected in the forthcoming years.

\section{Conclusions}
In this work we have provided a discussion on the possibility to detect and characterise DM through detection of a signal of circularly polarised photons. We showed that if DM is coupled to the SM by means of a parity violating interaction, a flux of highly polarised photons is expected from the GC of our galaxy. Prospects of detection of this signal in the near future by experiments like Fermi-LAT and e-ASTROGAM are not too optimistic, unless novel techniques are devised to reduce backgrounds and to improve the angular resolution and the sensitivity to the polarisation fraction. However, other sources such as cosmic accelerators could potentially lead to higher fluxes by means of an enhanced CR spectrum. We leave the study of these cases for a future work.

\section*{Acknowledgements}
The work of C.D. and M.C. was funded by the F.R.S.-FNRS through the MISU convention F.6001.19. The work of L.M. has received funding from the European Union’s Horizon 2020 research and innovation programme as part of the Marie Sklodowska-Curie Innovative Training Network MCnetITN3 (grant agreement no. 722104). Computational resources have been provided by the supercomputing facilities of the Universit\'e Catholique de Louvain (CISM/UCL) and the Consortium des \'Equipements de Calcul Intensif en F\'ed\'eration Wallonie Bruxelles (C\'ECI) funded by the Fond de la Recherche Scientifique de Belgique (F.R.S.-FNRS) under convention 2.5020.11 and by the Walloon Region.

\bibliographystyle{utphys}
\bibliography{refs.bib}
\end{document}